\documentclass{IEEEtran}
\usepackage{cite}
\usepackage{amsmath,amssymb,amsfonts}
\usepackage{graphicx}
\usepackage{textcomp,nicefrac}

\usepackage{microtype}
\DisableLigatures{encoding = *, family = *}
\usepackage{graphicx}
\usepackage{listings}
\usepackage{xcolor}
\usepackage{verbatim}
\def\BibTeX{{\rm B\kern-.05em{\sc i\kern-.025em b}\kern-.08em
T\kern-.1667em\lower.7ex\hbox{E}\kern-.125emX}}
\begin{document}
\title{HDPython: A High Level Python Based Object-Oriented HDL Framework  \quad (Feb 2021)}
\author{R. Peschke, K. Nishimura, G. Varner}

\maketitle
\newcommand{\pyhdl}{HDPython}
\newcommand{\argghdl}{HDPython}
\newcommand{\vhdl}{VHDL}
\newcommand{\oo}{object-oriented}
\newcommand{\Oo}{Object-oriented}
\newcommand{\fpl}{functional programming language}
\newcommand{\fpls}{functional programming languages}
\newcommand{\fsharp}{F-Sharp}
\newcommand{\hlpl}{high-level programming language}
\newcommand{\hlpls}{high-level programming languages}
\newcommand{\this}{this document}
\newcommand{\This}{This document}
\newcommand{\pl}{programming language}
\newcommand{\dataobj}{data-object}
\newcommand{\dataobjs}{data-objects}
\newcommand{\Dataobj}{Data-object}
\newcommand{\resph}{responsibility-handler}
\newcommand{\Resph}{Responsibility-handler}
\newcommand{\proc}{processor}
\newcommand{\Proc}{Processor}
\newcommand{\stdlogic}{std logic}
\newcommand{\stdlogicvector}{std logic vector}
\newcommand{\axistream}{AXI4-Stream}
\newcommand{\cpp}{C++}
\newcommand{\verilog}{Verilog}
\newcommand{\vin}{"in"}
\newcommand{\vout}{"out"}
\newcommand{\vinout}{in/out}
\newcommand{\ub}{UB}
\newcommand{\vsc}{Variable/Signal/Constant}

\newcommand{\necessary}{necessary}
\newcommand{\Necessary}{Necessary}
\newcommand{\getstreaminput}{\_get\_Stream\_input}
\newcommand{\getstreamouput}{\_get\_Stream\_output}

\newcommand{\portstreamsecondary}{pipeline\_in}
\newcommand{\portstreamprimary}{pipeline\_out}
\newcommand{\ventitylist}{entity\_list}

\newcommand{\vswitch}{v\_switch}
\newcommand{\vcase}{v\_case}
\newcommand{\codeblock}{Listing}
\newcommand{\Codeblock}{Listing}
\newcommand{\hdlconverter}{hdl\_converter}
\newcommand{\arggObject}{\argghdl\_Object}
\newcommand{\arggObjects}{\argghdl\_Objects}

\begin{abstract}
We present a High-Level Python-based Hardware Description Language (HDPython), It uses Python as its source language and converts it to standard VHDL. Compared to other approaches of building converters from a high-level programming language into a hardware description language, this new approach aims to maintain an object-oriented paradigm throughout the entire process. Instead of removing all the high-level features from Python to make it into an HDL, this approach goes the opposite way. It tries to show how certain features from a high-level language can be implemented in an HDL, providing the corresponding benefits of high-level programming for the user.
\end{abstract}


\section{Introduction}

\pyhdl\  is a library that allows the user to write Python  code and convert it to \vhdl\ (VHSIC-HDL, Very High Speed Integrated Circuit Hardware Description Language) \cite{VHDL}. It is licensed under the "MIT licence" and can be downloaded either via GitHub\cite{Github} or via PyPi \cite{PyPi}.  \This\ shows three of its most defining features. Firstly \argghdl\ is  fully object oriented, which allows the user to create high-level objects. This Object Oriented approach allows the users of the library to express their intentions in a cleaner manner, which can result in much cleaner code. Secondly, it allows the user to write very generic code. This reduces the amount of code duplication. Due to its  high-level templating mechanism, it can be used to write highly generic code. Thirdly, it uses a preexisting language, which allows the user to use the already well known and very powerful tools, which are provided in the Python language. Instead of having to learn both digital logic design practices AND a completely new language, the user only has to learn a few extensions provided by this library. 

\subsection{Competing Approaches }
The idea of generating firmware from a \hlpl\ is certainly not new; in fact some of these approaches have been around for decades.  There is a large number of approaches for High level Synthesis (HLS) and firmware generation (MyHDL \cite{myHDL}), but what sets \argghdl\ apart from the others is its high level approach to writing low level firmware. If compared to, for example, "VivadoHLS" \cite{HLS},  it is clear that "VivadoHLS" provides very strong support for generating algorithms out of existing c/c++ code and porting them over to firmware, but the code written in c/c++ has very little resemblance to the firmware generated. In fact most of the time the generated code will only be used as a black box IP core. For many applications this is a solid approach but in cases where one still wants to have the low level control over the actual implementation it is difficult to achieve that. Here \argghdl\ can really shine because it allows the user on one hand to write the firmware as low-level as is necessary to fulfill the task, and on the other hand gives a clear path of integrating high level features into the design. 
\argghdl\ is not an HLS tool, it is not designed to convert an arbitrary Python program into Register-transfer level (RTL). Its goal is to allow the user to describe RTL using a high level language. Further more its central design philosophy is to be an extensible framework which lets the user define its own means of conversion which allows the user to port high level language feature into designing RTL. Therefore it puts a big emphasis on exposing as much as possible its conversion mechanism.
Another popular code generation tool is "MyHDL". It allows the user to write firmware in Python. This is a task it does very well but in doing so it loses many of the features that make Python a \hlpl. Most noticeable here is classes. It provides rudimentary support for grouping signals together in containers but this is the highest level of abstraction that one can achieve.  With \argghdl\ classes are a core part of its design, furthermore \argghdl\ sees itself as a platform for the user to create their own abstractions on-top of the existing ones. As shown later in \this\ virtually every behavior of the library can be customised for individual classes. Lastly the approached that is most similarly to the approach presented in \this\ is spinalHDL \cite{spinalHDL}. Especially when it comes to 
2
 bundling signals together both approaches share some similarities. However \argghdl\ sets itself apart by being more \oo\ and putting the conversion front and center for the user to explore. In addition \argghdl\ is Python based, which is in the field it was developed for (High Energy physics), alongside with \cpp\ the standard language. According to the "TIOBE Index for January 2021" \cite{TIOBE} Python is the third most popular language while Scala (the base language of spinalHDL) is 34th making it one position more popular than \vhdl.

\section{\Oo\ Design}
\Oo\ Design has been the foundation of virtually all modern programming languages. Object-Oriented design has been proven excellent at hiding complexity. Humans are able to use extremely complicated technology, such as computer chips, since its complexity is hidden inside powerful abstractions, which reduce its complexity to just pushing buttons. 

\subsection{Programming in \vhdl\ without objects}
\begin{figure}
  \includegraphics[width=\linewidth]{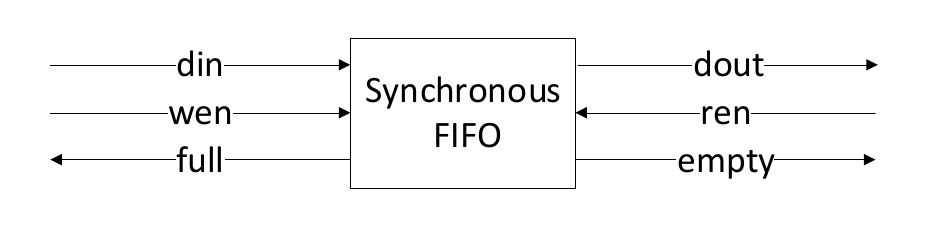}
  \caption{FIFO with native interface}
  \label{fig:NativeFIFO0}
\end{figure}

Listing \ref{lst:vhdl_FIFO} shows a typical example of an entity declaration in \vhdl. The entity declared on  \ref{lst:vhdl_FIFO} is a queue or "First in First out" (FIFO) module (see figure \ref{fig:NativeFIFO0}). A FIFO consists of two sides. The data input side and the data output side. In order to stay synchronized each side provides three signals. For the Data Input side these signals are: write data, write enable, and full. For the correct use of this FIFO, it is crucial to handle these signals precisely according to  specification. Yet when we look at the entity declaration the only thing that reveals which signals belong to which side of the FIFO is a  vague naming convention. \vhdl\ has no concept to describe these three signals belonging to one interface. A main reason for this is that each port has to be defined as an input or an output. Therefore it is not possible to make a record with these three signals. In this case the entire record and all its signals, would either be an input port or an output port.

\begin{figure}
  \includegraphics[width=\linewidth]{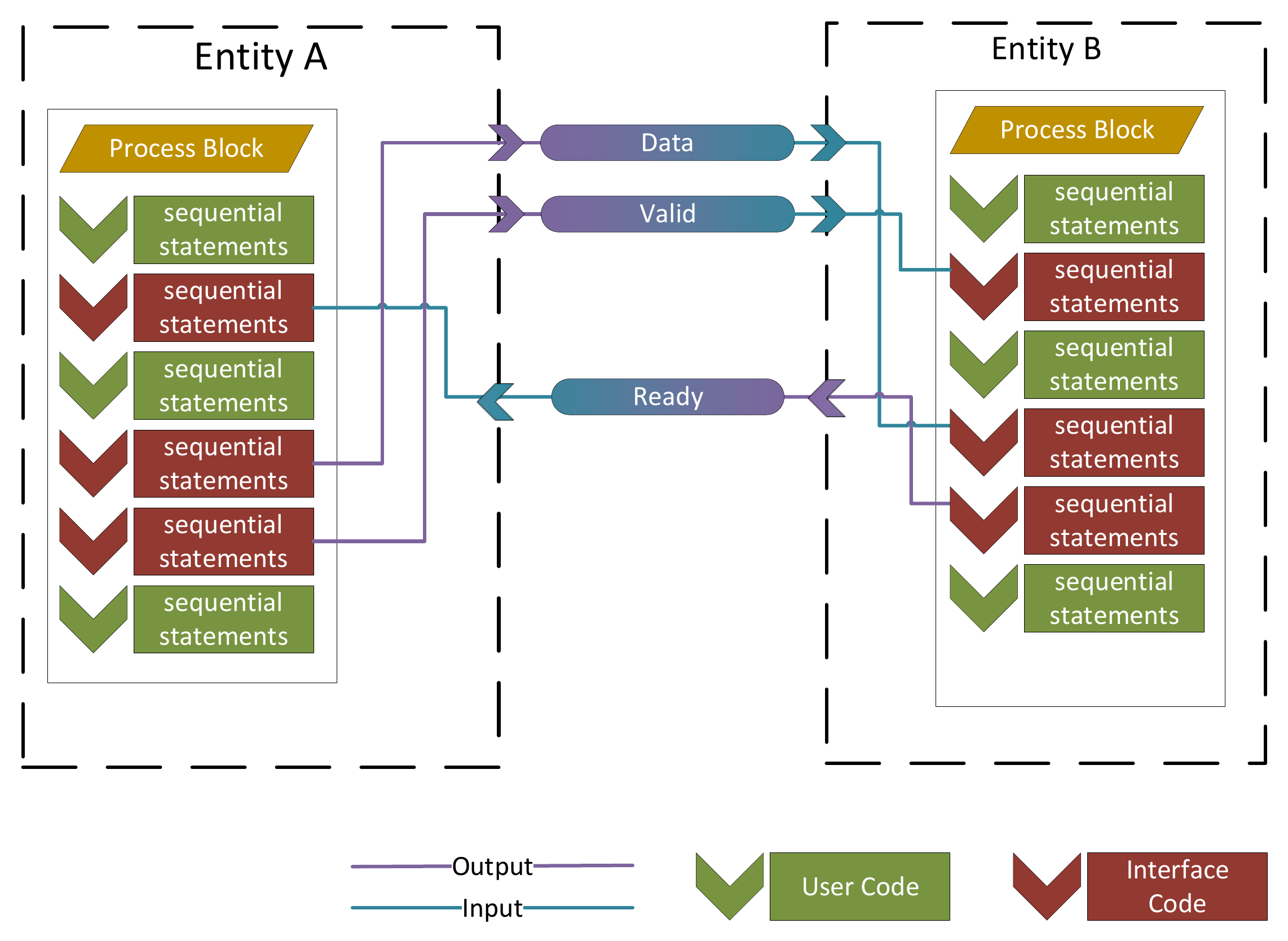}
  \caption{User Code and Interface code is intermingled. Interface code needs to be re-implemented for each entity. Interface code is hard to recognize. Inputs and outputs are independent objects. }
  \label{fig:PrimarySecondary0}
\end{figure}

Figure \ref{fig:PrimarySecondary0} shows, how two entities communicate with each other using three signals. As can be seen, it is not obvious that these three signals actually define one interface. The only thing that reveals this information is a  naming convention. Inside each entity, %
the signals are also handled  loosely. Their properties are usually checked/set at very different stages of the program. Therefore, the user code and the interface code are completely intermingled. This can get unmanageable quickly and leads to the typical "write only code", that is seen so often in  \vhdl.

\begin{lstlisting}[caption={
Typical declaration of a FIFO with a native interface \cite{nativFifo}. As shown in figure \ref{fig:NativeFIFO0}, the FIFO consists of three signals handling the writing of Data to the FIFO (din,wen, full) and three output signals (dout, ren, empty). The only hint that the user has on how to use this FIFO is a naming convention. But of course a naming convention does not communicate the timing behavior of the interface.  Compared the the more modern \axistream\ the native FIFO interface has the additional disadvantage of having two different interfaces for the data input side  and the data output side. 
},label={lst:vhdl_FIFO},captionpos=b,float,floatplacement=H]
entity FIFO_cc is
generic(
   DATA_WIDTH : natural := 16;
   DEPTH : natural := 5 
);

port(
   clk   : in  std_logic;
   rst   : in  std_logic;
   din   : in  std_logic_vector(
                DATA_WIDTH-1 downto 0); 
   wen   : in  std_logic;
   ren   : in  std_logic;
   dout  : out std_logic_vector(
                DATA_WIDTH-1 downto 0);
   full  : out std_logic;
   empty : out std_logic
);
end FIFO_cc;

\end{lstlisting}

\subsection{Programming in \argghdl\ with Objects}

To start with \argghdl\ a prior knowledge of any hardware design language is not required but, if possible, the names used to describe certain constructs are chosen to be similar to \vhdl, therefore user with a \vhdl\ background will find the overall structure familiar.  \codeblock\ \ref{lst:my_first_test_bench} shows a simple example of a counter. The Example starts by creating a class which inherits from "v\_entity" this marks the object as an entity object. V\_entities work very similar to to entities in \vhdl\ In our example the entity has no ports and defines only one function, the "architecture" function. Similarly to the architecture body of an entity in \vhdl\ this contains the internal structure of the entity. Inside the architecture function the first thing that we create is a clock generator. The clock generator is derived from  v\_entity object itself.  It has one (public) member which is the clock. Since it is a normal Python object the clock generated by the entity clock generator can be accessed like any other member of a Python class with "clkgen.clk". After this two signals are being created. One is the counter, the other one is the maximum value to which the counter should count. Both of these signals are of the type "v\_slv(32)", which translates to a 32 bit standard logic vector. The second argument in "max\_cnt = v\_slv(32,300)" is the default value. Since this is ordinary Python code the initial value could have been given in hexadecimal as well. The default initial value is zero. The next construct is a function definition "def proc():" which is modified by a decorator "@rising\_edge(clkgen.clk)". This special decorator takes the function that is generated here and appends it to the "on\_change" list of the signal it received as an argument "clkgen.clk". But instead of executing on ever signal change this decorator makes sure that it is only executed on a rising edge. Once the simulation has started the function "proc" gets executed on every rising edge of the signal "clkgen.clk". The function proc does three things. First it increments the counter signal by one. For this it uses the stream (bit-shift) operator. The idea is that the new value gets streamed into the existing and persistent object. Since the object might be connected to another object it is important that there is no new object generated but the existing object is modified. Similar to \cpp, \argghdl\ uses the stream operator to stream data into objects (signals). In addition to the left shift operator (stream in operator) there is a right shift operator (stream out operator). Future examples will show them in action. Next, the counter is compared to max\_cnt and if it is bigger or equal it is then set to zero again. The last statement is "end\_architecture()" which handles some book keeping such as giving all the signals the correct name etc. It needs to be executed at the end of the "architecture" block. 

On the use of streamer (bit shift) operators: In other programming languages, such as \cpp\, overloading the streamer (bit-shift) operator to send data into an object (or read from it) has been part of the language for decades. It is a clear way of communicating to the user that this object is not replaced with a new object but its content is modified. Of course it would be possible to use a "next" function (or property) to achieve the same behavior. In the end it is a matter of taste which style one prefers but overall using the streamer (bit-shift) operator gives a very natural indication of data flow. 

On variables and signals: \argghdl\ allows you to define signals as either a variable or a signal. However since this "variableness" (blocking value assignment) and "signalness" (non-blocking value assignment) is part of the object definition there is no need for two separate assignment operators. The streamer (bit-shift) operator will work on both types.

\begin{lstlisting}[caption={ 
my\_first\_test\_bench
},label={lst:my_first_test_bench},captionpos=b,float,floatplacement=H]
class my_first_test_bench(v_entity):
    def __init__(self):
        super().__init__()
        self.architecture()

    @architecture
    def architecture(self):
        clkgen = clk_generator()
        counter = v_slv(32)
        max_cnt = v_slv(32,300)

        @rising_edge(clkgen.clk)
        def proc():
            counter << counter + 1
            if counter >= max_cnt:
                counter << 0 
        end_architecture()
\end{lstlisting}

\begin{figure}
  \includegraphics[width=\linewidth]{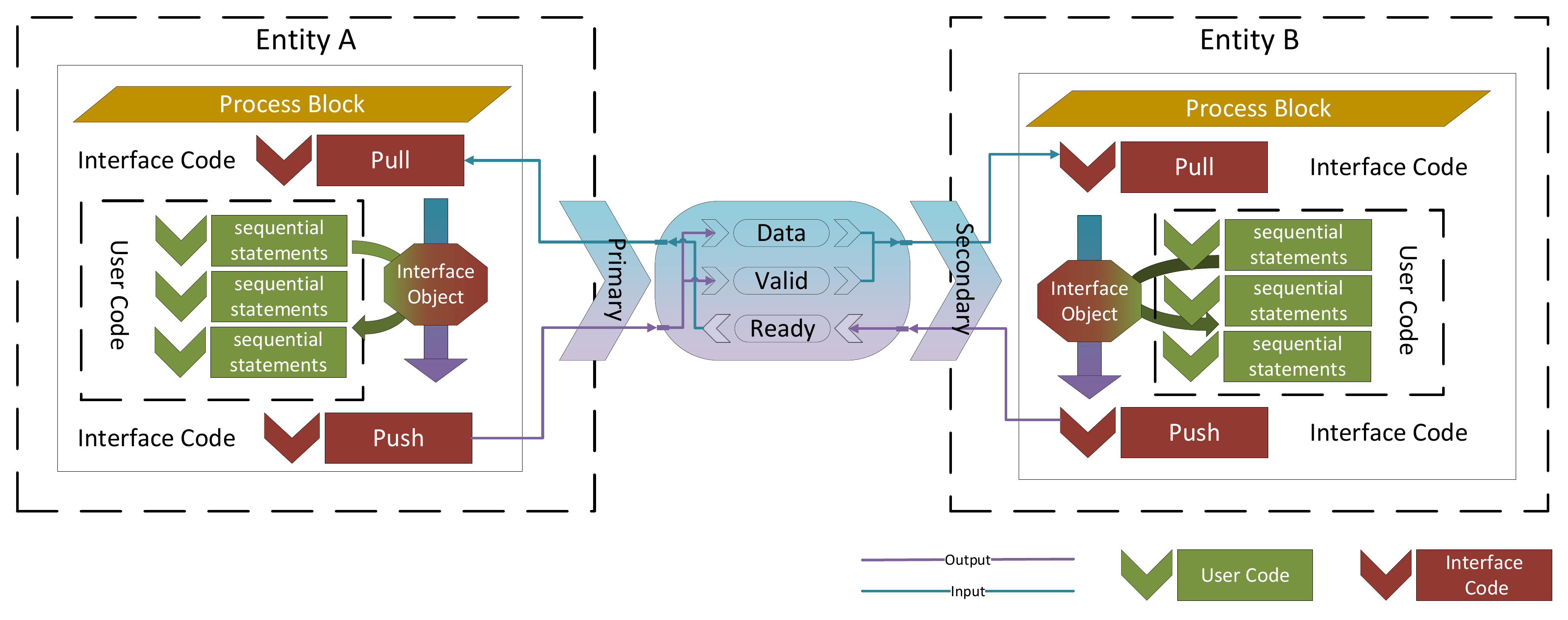}
  \caption{User Code and Interface code are clearly separated. This allows for the reusing of well known (tested) functions. Interface code is immediately recognizable}
  \label{fig:PrimarySecondary_cl0}
\end{figure}

Communication between Entities with Classes and Application Programming Interfaces (API). In \argghdl\ the interface is defined as an object (see figure \ref{fig:PrimarySecondary_cl0}). The interface object contains all three signals as well as their directionality. Each entity has one port for the entire interface. This way it is clear that these signals belong to one interface. Instead of defining a port as either input or output, the port are defined as either primary or secondary. In a primary port, the signals have the directionality defined in the interface class. In a secondary port, the directionality is flipped. 

Inside the entity the interface is handled by a handler class. This class contains the API for interacting with the interface. Thereby, the user code and the interface code are separated. This makes it easier to write clean code that is easily understood. In addition, since the user never directly interacts with the interface signals, the interface author can ensure that the interface is used correctly. 

\subsubsection{Example: Clocked Entity sending data via \axistream\ link}
\Codeblock\ \ref{lst:argg_data_source} shows an example of a data source entity. In this case the entity just counts up numbers. For interaction with other entities, it uses the \axistream\ interface \cite{axiFIVO}. Since it is a data source, it uses the "port primary" version of the interface. On the architecture level it first defines a handler object called "data out", which handles the interaction with the interface. The handler is used inside the process block. The process block is executed on every rising edge of the clock. Inside the process block the handler is first checked with the "if statement". Since the handler class is a high level object the author has the possibility to overload the "bool" function, which is called inside the test block of the "if" statement. 

This allows for expressive code. If the bool function of the interface handler returns true, the interface is ready to use. On the next line, the data can be assigned to the handler object, which is equivalent to calling the send data function. The user of the library never needs to know the internal structure of this interface nor does the user need to know the protocol for sending data. The users can always be sure that, as long as they are only using the functions provided by the API, the interface will always work correctly. 

\subsubsection{Example: Interface Class}
How the interface class works is now described. Shown in \codeblock\ \ref{lst:argg_FIFO_trans} is the interface class for an \axistream\ interface with a data width of 32 bits. It defines all signals it needs to operate. It sends from primary to secondary the following signals: valid, last and data. From secondary to primary it sends the ready signal. Primary and secondary are defined relative to the data flow. The data flows from primary to secondary.

\subsubsection{Example: Interface Handler Class}

\Codeblock\ \ref{lst:argg_FIFO_primary} describes how the interface handler class works, the interface class has to inherit from an \argghdl\ base class called "v\_class\_primary". The constructor takes one argument, which is the interface class, then it  defines a variable port of that type and connects the two objects together. Note that since these two object, "axi\_out" and "TX", are objects they know exactly which signal has to go in which direction in order to make the interface work. The handler class defines three additional functions. First is the send data function. It takes any data it receives and sends it out through the interface class. In addition, it also sets the steering signal "valid" to high. The second function is the "ready\_to\_send" function. It is used to check if the interface is ready to send new data. The interface is only ready to send data if it is not already sending data. Therefore, the "valid" signal needs to be low in order to send more data. The third function is the "\_on\_Pull" function. This function is called on every clock cycle at the beginning of the process block. In this case, it checks if the "ready" signal send from the secondary is high. If yes, it means that the secondary has read the current data and is ready to receive new data. If the ready signal is high, the steering signals can be reset and new data can be sent.

\subsubsection{Example: Entities are Objects}

Entities are themselves objects, therefore each port of an entity can be accessed as a member variable of this object. The example in \codeblock\ \ref{lst:argg_entities_are_classes} shows how different entities can be connected together. Inside the architecture block of the entity, we first create a clock generator entity. The instance is called "clkgen". The instance has an output port "clk", which can be accessed like any other member variable in Python. We use  the clock provided by "clkgen" in the constructor for the other entities. Now that we have created the entity "cnt" (short for  counter) and "axiPrint" we can connect its signals together. We do this by simply assigning the output of the counter to the input of the "axi print" object. Since the interface is an object itself, it knows exactly which signals have to go in which direction. The signals do not need to be repeated on the test bench entity. Everything is contained inside the entities themselves.

\begin{lstlisting}[caption={Instead of defining individual inputs and outputs, an interface object is defined. The object "axiStream\_32" contains all information about input and output signals. The object brings a handler with it. The handler is used for communication with the interface. The handler provides the API for the interface. The user never directly interacts with the data members.
},label={lst:argg_data_source},captionpos=b,float,floatplacement=H]
from HDPython import *
from HDPython.examples import *
class Counter(v_clk_entity):
  def __init__(self, clk ):
    super().__init__(clk)
    self.Dout = port_out(axiStream_32())
    self.architecture()

  @architecture
  def architecture(self):
    data = v_slv(32)
    data_out = get_handle(self.Dout)
    
    @rising_edge(self.clk)
    def proc():
      if data_out:
        data_out << data
        data << data + 1
        
\end{lstlisting}

\begin{lstlisting}[caption={This class describes the signals required for the interface. In addition to just storing the type of the data, it also stores the direction of the data. By definition, data flows from primary to secondary. port\_out defines a signal that goes from primary to secondary. port\_in defines a signal that goes from secondary to primary},label={lst:argg_FIFO_trans},captionpos=b,float,floatplacement=H]
class axiStream_32(v_class_trans):
  def __init__(self):
    super().__init__()
    self.valid = port_out( v_sl()    )
    self.data  = port_out( v_slv(32) )
    self.ready = port_in ( v_sl()    )
\end{lstlisting}

\begin{lstlisting}[caption={This class handles the primary(sender) side of the interface. The constructor takes the Interface class as argument. It makes a local copy in a variable. It connects the local copy to the Input argument. The object knows exactly which signals have to go in which direction. The \_onPull functions allows the creator of the interface class to inject functionality on every clock cycle (before the user's code)},label={lst:argg_FIFO_primary},captionpos=b,float,floatplacement=H]
class axiStream_primary(v_class_primary):
  def __init__(self, Axi_Out):
    super().__init__()
    self.tx =  variable_port_out( Axi_Out)
    Axi_Out  << self.tx
 
  def send_data(self, dataIn ):
    self.tx.valid   << 1
    self.tx.data    << dataIn    
    
  def ready_to_send(self):
    return not self.tx.valid
    
  def _onPull(self):
    if self.tx.ready: 
      self.tx.valid << 0 
      self.tx.last  << 0 
\end{lstlisting}

\begin{lstlisting}[caption={AxiPrint is an entity that prints out the values of the stream. It uses the same interface class as "Counter" but since it wants to consume the data it is a secondary port. In order to connect "Counter" with "AxiPrint" the Data\_in / Data\_out Member needs to be connected. Since the interface class knows which signal goes in which direction the signals can just assigned to each other.},label={lst:argg_entities_are_classes},captionpos=b,float,floatplacement=H]
class tb(v_entity):
  def __init__(self):
    super().__init__()

  @architecture
  def architecture(self):
    clkgen  = clk_generator()
    cnt     = Counter(clkgen.clk)
    axPrint = AxiPrint(clkgen.clk)
    axPrint.D_in << cnt.Dout
    end_architecture()

class AxiPrint(v_clk_entity):
  def __init__(self, clk ):
    super().__init__(clk)
    self.D_in = port_in(axiStream_32())

\end{lstlisting}




\subsection{Classes: Combination of Data and Functions}

Let us have a deeper look into classes. Classes are usually thought of as the combination of data and functions. That means they can contain both member data and member functions. Hardware Description Languages (HDL), usually deal with many different types of data, and functions. For example, there are variables, signals and ports. In addition to normal functions, HDLs have to deal with functions, procedures, processes, and entities. In order for an object to be a fully high-level object, it is mandatory that classes in \argghdl\ are able to contain all these different parts of the language.

\begin{figure}
  \includegraphics[width=\linewidth]{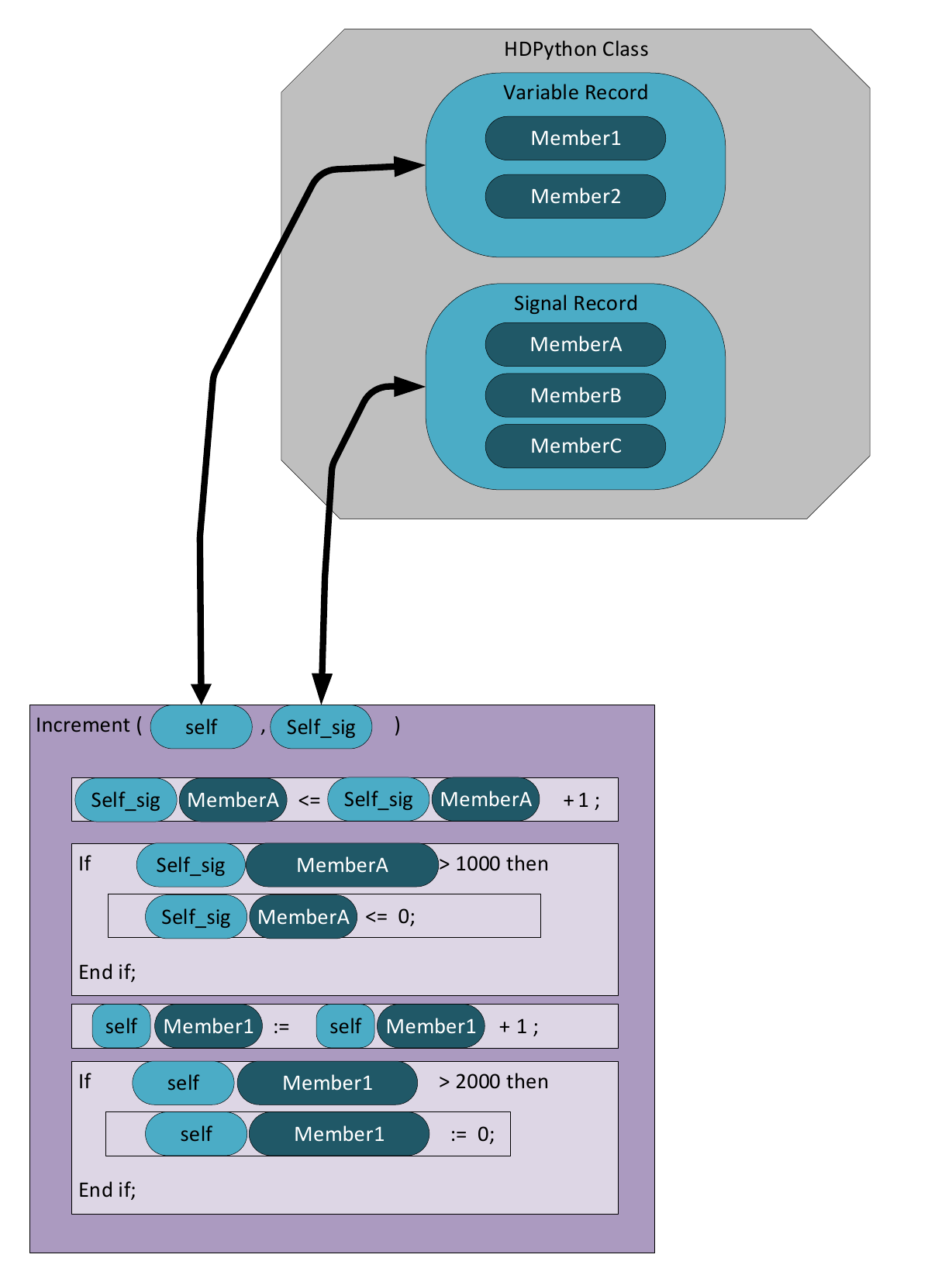}
  \caption{Native FIFO interface: requires enable to be set to '0' immediately when empty goes to '1'. This needs to be done using combinatorial logic.}
  \label{fig:native_FIFO_handler}
\end{figure}

Classes can contain variables, signals, and combinatorial logic (asynchronous push/pull functions) are useful for defining interfaces to other entities. In the example, shown in figure \ref{fig:native_FIFO_handler}, we want to access a FIFO with a native interface. This specific FIFO requires the enable signal to be always zero if the empty signal is high. It needs to be done immediately and cannot wait for the next clock cycle; therefore, it needs to be done asynchronously. In \argghdl\ it is possible to encapsulate all this complexity inside the interface handler class. The user of the class does not need to know that there is any additional logic. In fact, if the author of the classes uses the same public API, the user can switch back and forth between both types of interfaces without having to change the code.

\begin{figure}
  \includegraphics[width=\linewidth]{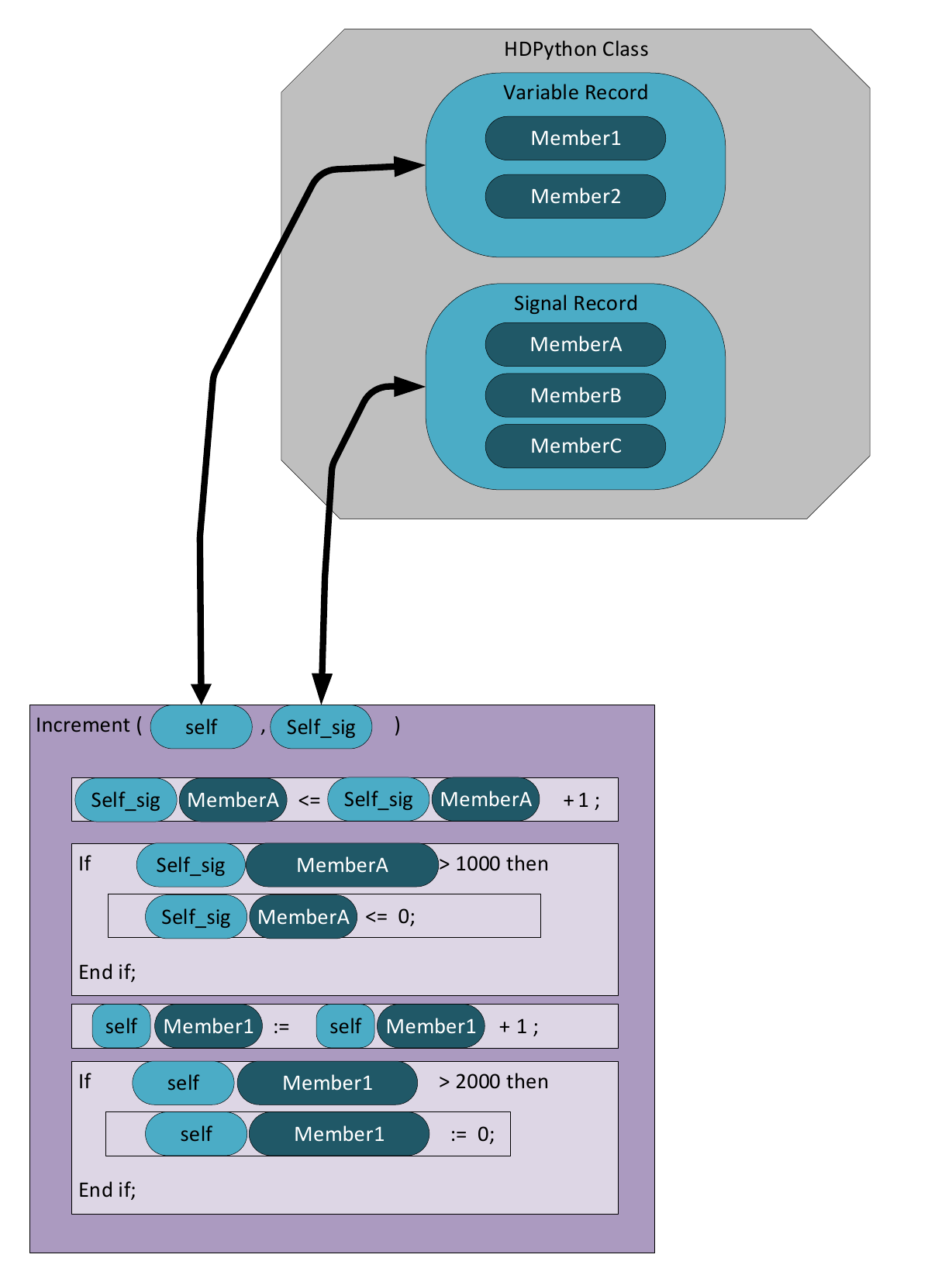}
  \caption{Translation. The \argghdl\ Class gets separated into its variables and signals. Since handler classes are always associated to exactly one process block the variables and signal records are given to the procedure with read and write access (inout).}
  \label{fig:signal_variable_class}
\end{figure}

As shown in figure \ref{fig:signal_variable_class}, classes in \argghdl\ are translated into \vhdl\ by first dividing its members into signals or variables. It then creates a record for each type. In the next step, the converter translates the member functions. Similarly to Python, the first argument is the object itself. Since the object now consists of two records, the function/procedure gets both of these records as arguments. Since handler classes are associated with only one process block. each function/procedure gets read/write (inout) access to both the signal record and the variable record. Only interface classes (and primitive data types such as \stdlogicvector) can be used by more then one process block.   

\section{Generic Programming}

The next feature any language needs in order to be considered a high-level language is generic programming. Even though \vhdl\ provides a limited amount of generic programming it falls short when compared to a high-level language. Given the dynamic nature of Python it naturally provides possibilities for very generic programming. \argghdl\ tries to preserve the generic programming model from Python as much as possible. Especially for  interface classes this is advantages. \Codeblock\ \ref{lst:argg_Interface_Classes_generic} shows the actual implementation of the interface class for an \axistream\ interface. The data type is an argument to the constructor of the class. The data type can be every plain type, that includes standard logic vectors of different sizes, integers etc. In addition, it can also be records and arrays. For every new data type, the templating mechanism will create a new class. 


\begin{lstlisting}[caption={For the actual implementation of the \axistream\ interface class the data object is a template variable. For each new data type the templating machine will create a new class. Data type can also be records and arrays.},label={lst:argg_Interface_Classes_generic},captionpos=b,float,floatplacement=H]
class axiStream(v_class_trans):
  def __init__(self,Axitype):
    super().__init__()
    self.valid  = port_out( v_sl()  )
    self.last   = port_out( v_sl()  )
    self.data   = port_out( Axitype )
    self.ready  = port_in ( v_sl()  )

\end{lstlisting}

\section{Conversion to \vhdl}

The conversion from Python to \vhdl\ is separated into different steps. In the first step the entity that should be converted needs to be instantiated. During the instantiating process all  the connections between the individual signals entities and classes are  created. At this point the entity would be able to run as a simulation in Python. The entities and all sub-entities have been instantiated at least once. Only objects that are currently instantiated get translated to \vhdl. The library has a shadow register which contains all \argghdl\ objects that are created during one session. In the next step it starts looping over all objects that have been created. 

In order to allow the user of the library a more granular control over the conversion to \vhdl\ every part of the conversion can be modified. In the end, the Python code and the \vhdl\ code can be made completely independent of each other. 

The Conversion is subdivided into three stages Firstly parsing of the Python source files. This is done with a standard Python library called AST. It is part of the reference implementation of CPython. Second, the \argghdl\ visitors, in order to traverse the AST, \argghdl\ has its own visitor class. This class does not only traverses through the syntax tree but also carries all the context information. For example, it contains a list of all the functions and variables that are defined in this context. In addition, it stores information about the environment, for example if the code that is currently looked at is a process, function, procedure, etc. It then makes the decision which function to call from the \argghdl\ object converter. It is important to note that the visitor itself does not make any conversion it just calls the function from the \argghdl\ converters. Last is the \argghdl\ object converters. Each object possesses its own converter, which defines how it is supposed to be converted to \vhdl. This converter also defines how each interaction with the object is converted. This gives the user a very granular way of defining how the object will converted to \vhdl. In addition, its modular design makes it very easy to replace one set of object converters for another set, which would allow a conversion to, for example, \verilog\ or even \cpp.



\section{Conclusions}

The goal of \this\ was to show an object oriented approach towards  Register-transfer level design using an well established programming language. The benefits of this approach have grouped into three chapters. In chapter III the benefits of being able to group signals into objects have been demonstrated. chapter IV shows the how the generic programming model of Python can be preserved and chapter V lists the benefits of using a existing and well established language. Lastly chapter VI is giving a very prev introduction into the conversion mechanism from Python to \vhdl. Even though a complete description of the conversion process would be extensive, a climes was giving in the mechanism and its defining feature which is before all others its expandability. Already in its early stage the \argghdl\ library has shown the benefits an object oriented approach to RTL design brings. 


%

\begin{thebibliography}{}


\bibitem[1]{Github} “HDPython/hdpython.” GitHub,  2021-02-05, 
https://github.com/HDPython/HDPython

\bibitem[2]{PyPi} "project/HDPython/" PyPi, 2021-02-05, 
https://pypi.org/project/HDPython/

\bibitem[3]{HMB}
L.~Ruckman, G.~Varner and K.~Nishimura,
Nucl. Instrum. Meth. A \textbf{623}, 365-367 (2010)
doi:10.1016/j.nima.2010.02.250


\bibitem[4]{KLM} "The TARGETX ASIC for the Belle II Muon Detector Scintillator Upgrade,"
G. Varner, B. Edralin, I. Mostafenezhad, X. Shi, presented at the Nov.
2015 IEEE Nuclear Science Symposium, Nov. 6, 2015, San Diego, CA.

\bibitem[5]{Python} G. van Rossum 1995,
      Python tutorial,
      Centrum voor Wiskunde en Informatica (CWI)
      (Amsterdam) 

 \bibitem[6]{VHDL} "IEEE Standard VHDL Language Reference Manual," 
 in IEEE Std 1076-1987 , vol., no., pp.1-218, 31 March 1988, 
 doi: 10.1109/IEEESTD.1988.122645.
 
 \bibitem[7]{HLS}  Vivado Design Suite User Guide(High-Level Synthesis),
   UG902 (v2019.2) January 13, 2020
 
 
 \bibitem[8]{myHDL} FPGA '15: Proceedings of the 2015 ACM/SIGDA International Symposium on Field-Programmable Gate Arrays February 2015 Pages 28 - 31 https://doi.org/10.1145/2684746.2689092
 
\bibitem[9]{spinalHDL} Papon, C. (2018). SpinalHDL. FOSDEM VZW. https://doi.org/10.5446/42349 
\bibitem[10]{TIOBE} "TIOBE Index for January 2021", TIOBE, 2021-02-05, https://www.tiobe.com/tiobe-index/  
 \bibitem[11]{nativFifo} Embedded FIFO Generator v1.0 Logi CORE IP Product Guide PG327 (v1.0) July 14, 2020

 \bibitem[12]{axiFIVO} AXI4-Stream FIFO v4.1LogiCORE IP Product GuideVivado Design SuitePG080 April 6, 2016

\end{thebibliography}
%

\end{document}